\documentclass{sigchi}

\usepackage{balance}  
\usepackage{graphics} 
\usepackage{times}    
\usepackage{url}      
\usepackage{enumitem}

\makeatletter
\def\url@leostyle{%
  \@ifundefined{selectfont}{\def\UrlFont{\sf}}{\def\UrlFont{\small\bf\ttfamily}}}
\makeatother
\def\pprw{8.5in}
\def\pprh{11in}

\usepackage[pdftex]{hyperref}
\hypersetup{
pdftitle={SIGCHI Conference Proceedings Format},
pdfauthor={LaTeX},
pdfkeywords={SIGCHI, proceedings, archival format},
bookmarksnumbered,
pdfstartview={FitH},
colorlinks,
citecolor=black,
filecolor=black,
linkcolor=black,
urlcolor=black,
breaklinks=true,
}

\makeatletter
\def\@copyrightspace{\relax}
\makeatother

\begin{document}

\title{MetaSpace II: Object and full-body tracking for interaction and navigation in social VR}

\numberofauthors{2}
\author{
  \alignauthor Misha Sra\\
    \affaddr{MIT Media Lab}\\
    \affaddr{Cambridge, MA, 02142 USA}\\
  \alignauthor Chris Schmandt\\
    \affaddr{MIT Media Lab}\\
    \affaddr{Cambridge, MA, 02142 USA}\\
}

\teaser{
\centering
\includegraphics[width=1.0\textwidth, height=0.25\textwidth, keepaspectratio]{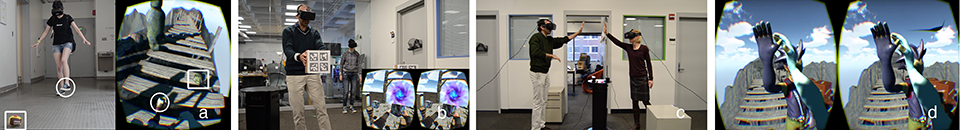}
\caption{MetaSpace II is a social Virtual Reality system where multiple users can not only a) see and hear but also walk around in space, and b) grasp and manipulate objects. When users touch or pick up an object in the virtual world, they simultaneously also touch or pick up it's counterpart in the physical world. (c) In MetaSpace users interact with each other both in the real, and d) the virtual world.}
\label{fig:teaser}
}

\maketitle

\begin{abstract}

MetaSpace II (MS2) is a social Virtual Reality (VR) system where multiple users can not only see and hear but also interact with each other, grasp and manipulate objects, walk around in space, and get tactile feedback. MS2 allows walking in physical space by tracking each user's skeleton in real-time and allows users to feel by employing passive haptics i.e., when users touch or manipulate an object in the virtual world, they simultaneously also touch or manipulate a corresponding object in the physical world. To enable these elements in VR, MS2 creates a correspondence in spatial layout and object placement by building the virtual world on top of a 3D scan of the real world. Through the association between the real and virtual world, users are able to walk freely while wearing a head-mounted device, avoid obstacles like walls and furniture, and interact with people and objects. Most current virtual reality (VR) environments are designed for a single user experience where interactions with virtual objects are mediated by hand-held input devices or hand gestures. Additionally, users are only shown a representation of their hands in VR floating in front of the camera as seen from a first person perspective. We believe, representing each user as a full-body avatar that is controlled by natural movements of the person in the real world (see Figure 1d), can greatly enhance believability and a user's sense immersion in VR.

\end{abstract}

\vspace{-1em}
\category{H.5.1.}{Information Interfaces and Presentation (e.g. HCI)}{Multimedia Information Systems}: {Artificial, augmented, and virtual realities }

\section{Introduction}

First person experiences of the real world represent a standard to which all mediated experiences are compared, either mindfully or otherwise \cite{steuer1995defining}. Bodily immersion in VR is rooted in the way the body is able to redirect a perception of itself as an object into virtual space (e.g. proprioception) and through this mirror image, the familiar body is also made the embodied subject during interaction \cite{sageng2012philosophy}. Presence in VR is based on the perception of input through visual, auditory, and kinesthetic senses. For enhancing presence, we need to incorporate the participant as a part of VR such that, head movements result in motion parallax, locomotion results in translation in space, proprioceptive cues are incorporated, vestibular responses are stimulated, feedback is provided through multiple sensory channels, and the user has agency. 

Ever since its conception in the 1960's, head-mounted virtual reality systems have been primarily concerned with the user's visual senses \cite{sutherland1968head} followed by spatial audio \cite{begault19943}.  Since two major problems faced by virtual environment designers have been lack of haptic feedback and constraints imposed on physical movement \cite{kohli2005combining}, we attempt to address them in our prototype system. Additionally, despite new HMD characteristics like low latency head tracking, wide field of view and custom input/output devices, current VR experiences lack proprioceptive cues related to locomotion which could limit a user's sense of presence in the virtual environment. In most existing VR systems, a seated user is presented with two contradictory motion cues; the visual stimulus signaling movement to the brain, while the vestibular system indicating a lack of movement creating a disconnect that can be a bit jarring. These conflicting inputs can also produce motion sickness and postural instability \cite{kennedy1996postural} resulting in, possibly believable, but sub-optimal immersive experiences. For exploring VR environments, walking in place is also not considered sufficient as it lacks the proprioceptive cues of actual walking \cite{williams2007exploring}. We propose a social virtual reality system that incorporates body awareness, physical presence, spatial understanding, sense of orientation, object manipulation, communication, and multi-sensory feedback to overcome some of the limitations of existing systems. Along with other researchers, we also believe including these features can greatly enhance the user's sense of presence and immersion in VR \cite{brooks1999s}.

In this paper we present an immersive and interactive social VR system with possible applications in domains such as entertainment, therapy, travel, real estate, education, social interaction and professional assistance. We use low-cost Kinect (RGB-D) devices to track multiple users and objects (positions and orientations) in real-time and use head-mounted displays (Oculus Rift DK2) for tracking head rotations and providing visual output. The main research contributions of our work are the following:
\begin{enumerate}[label= --]
\item We build our virtual environment by 3D scanning the physical environment. This 1:1 mapping allows users to walk freely in space without fear of colliding with walls or furniture. Their focus stays on the virtual world and maintains their sense of believability and immersion. This can lead to a smoother and uninterrupted experience which can in turn bring forth an enhanced sense of enjoyment.
\item The scanned physical environment is textured to make it visually different from reality. This helps create a sense of being somewhere else and allows us to use the same physical space to create many different virtual experiences.
\item Given that one of the key features of a VR system is interaction, gaining an understanding of the specific ways in which they facilitate this interaction is important. We use full-body avatars that are directly controlled by the movements of the user in the real world for enhancing engagement in our social VR system. 
\item We use skeletal tracking that allows Kinect to recognize people and follow their actions. This, in turn, enables users to walk naturally in the real world, providing all proprioceptive cues related to locomotion in the virtual world, and thus positively impacts the user's sense of presence in VR. Additionally it also provides a more intuitive and body-centric way of interacting with objects in VR.
\item We use the Kinect to track objects in the real world that have virtual counterparts. This provides users haptic feedback which has been shown to both enhance immersion and also make virtual tasks easier to accomplish in VR.
\item As it is a multi-person system, users can not only see each other, but also observe the consequences of each other's actions reflected through the actions of their avatars and their interactions with virtual objects. For instance, if a user grabs a virtual cube (that has a cardboard box physical counterpart) and puts it on top of another virtual cube, each user will perceive the transformation (i.e., orientation and position changes) of the cubes from their avatar's point of view. 
\item Our system is entirely based on low cost hardware and freely available software platforms and tools.  
\end{enumerate}

A major challenge for building a social VR world is performance in real-time. In our system, most of the computational load is concentrated in real-time object and user tracking followed by network transmission of the data and rendering of the visual output for each user. It was, therefore, crucial to make the system computationally tractable for object and user tracking, fast rendering, and data transportation. We achieved this through various strategies like optimizing the computer vision pipeline for object tracking on the server, using a connectionless transmission model for data communication, caching, and optimizing model geometry, lighting, textures and shaders at the client end.

\section{Related Work}

The visual medium evolved from early cave art to the realistic paintings of the classical era to photographs and videos. Artists have always been inspired by the real world and viewers have traditionally experienced a work of art from the artist's point of view. In 1968, Sutherland introduced the notion of the viewer being able to choose their point of view, an idea that forms the basis of VR as we know it today  \cite{sutherland1965ultimate}. Virtualized Reality \cite{kanade1995virtualized} was an realization of the user choosing a viewing angle while looking at a digital scene through a stereo-viewing system and it was implemented by using techniques from computer vision and computer graphics. While VR environments are typically constructed using CAD models, Virtualized Reality starts with the real world and virtualizes it. Another related concept is Substitutional Reality that involves a modification of normal perceived reality. In an experiment users viewed a live video feed interspersed with previously recorded video through an HMD and were unable to discriminate between the two \cite{suzuki2012substitutional}. As perception, presence, and immersion are all intertwined in VR, our goal is to create experiences that are perceptually believable, impacting the user's sense of presence through an immersive, though not necessarily realistic, system. In the remainder of this section, we describe other research themes that are incorporated in our prototype design.

\subsection{Passive Haptics}

The majority of research on interaction techniques for VR has focused on gesture and voice input as prime ways of interacting in VR. These interfaces build on users' pre-existing knowledge of the everyday life and experiences \cite{jacob2008reality}.  Ideally, we would be able to interact with VR just like we interact with the real world. In addition to input techniques, researchers have also pursued a few different approaches for providing haptic feedback e.g. motion platforms \cite{stewart1965platform}, exoskeletons \cite{hale2014handbook}, and vibrotactile gloves \cite{sturman1994survey}. While these approaches have been successful at giving users the experience of walking, providing some limited haptic feedback, they are not well suited for recreating the experience of grasping objects or touching a wall \cite{cheng2015turkdeck}. Opportunistic Controls are a class of user interaction techniques for augmented reality (AR) that leverage physical object affordances to provide passive haptics to ease gesture input, simplify gesture recognition, and provide tangible feedback to the user \cite{henderson2008opportunistic}. In AR, the user can see the real world objects which they are using to control or interact with digital elements. However, that is not the case in VR and we think this brings forth new opportunities for transforming mundane physical objects into magical virtual items for creating immersive VR experiences.

An early example of the use of a real object associated to a similar but not identical virtual object was in a desktop VR environment, where a doll head model was used to control a brain visualization \cite{hinckley1994passive}. Passive haptics have been shown to both enhance immersion in VR and also make virtual tasks easier to accomplish by providing haptic feedback. For example, Hoffman found adding representations of real objects, that can be touched, to immersive virtual environments enhanced the feeling of presence in those environments \cite{hoffman1998physically}. Lindeman found that physical constraints provided by a real object could significantly improve performance in an immersive virtual manipulation task \cite{lindeman1999towards}. For example, the presence of a real tablet and a pen enabled users to easily enter virtual handwritten commands and annotations \cite{poupyrev1998virtual}.  

In another study, participants had to place a book on a chair at the opposite end of a room by walking across a ledge over a deep pit. The researchers added a real wooden plank for users to walk on, that corresponded with a virtual ledge, and compared the differences in response when the ledge was only virtual. Since participants could feel a height difference between the wooden plank and the floor below, it enhanced the illusion of standing on the edge of a pit. Results showed significant difference in behavioral presence, heart rate, and skin conductivity changes in the wooden plank group of users \cite{insko2001passive}. Passive haptics have also been used in VR therapy. In one study, a physical spider toy replica was present in the virtual environment so that when the patient's virtual hand reached to touch the virtual spider, she could feel the furry texture of the toy spider \cite{carlin1997virtual}.

\subsection{Body, Environment and Social Awareness} 
Body awareness refers to the familiarity and understanding that people have of their own bodies, for example, awareness of the relative position their body parts (proprioception) and a sense of their range of motion \cite{jacob2008reality}.  The brain integrates information from proprioception and from the vestibular system into its overall sense of body position, movement, and acceleration \cite{wiki:proprioception}. Based on these innate abilities, research in VR interfaces has explored a variety of input techniques though much of VR interaction is dominated by hand-based input systems. Our prototype goes beyond hand-based input by allowing users to experience and interact in VR using their full body, similar to how they experience the real world. By representing users' bodies in the virtual world, our system allows users to perform tasks relative to the body (egocentric), enhancing their sense of presence.

In the real world, people have a physical presence in their spatial environment. We are surrounded by objects and landscape that facilitate our sense of orientation and spatial understanding \cite{jacob2008reality}. For example, the horizon gives us a sense of directional information, occlusions give us relative distance cues while atmospheric color, fog, lighting, and shadow provide depth cues \cite{bowman20043d}.  People also develop skills to manipulate objects in their environment, such as picking up, positioning, altering, and arranging objects \cite{jacob2008reality}. Our VR system uses these skills along with people's innate sense of motion and proprioception for interacting naturally with the virtual world, grasping and manipulating objects in both the real and virtual environments. 

As humans, we are generally aware of the presence of others and since childhood are taught skills for social interaction. These include verbal and non-verbal communication, the ability to exchange physical objects, and the ability to work with others to collaborate on a task \cite{jacob2008reality}. Our system uses social awareness and skills by representing users's presence as full-body avatars and by making the avatars' actions visible. Additionally the environment allows for multiple co-located users to interact with each other both in the real and virtual worlds and collaborate on tasks.

\subsection{Perception, Immersion, and the VR Experience}

Husserl says experience is characterized by intentionality and Heidegger calls our most basic experiential relation to the world is a practical one. According to Merleau-Ponty, experience is essentially corporeal. They all agree that a human is not a ``passive" observer responding to stimuli, but an active and creative force shaping the stimuli of their experience \cite{klevjer2012enter}. Gibson adds that reality of experience is grounded in action and the construct of affordance presents a new fundamental basis for reality in the relationship between actor and environment \cite{sageng2012philosophy}. These ideas lead to the notion of the body being the medium for the perception of space \cite{merleau1996phenomenology} and in a perfect VR experience, we would not be able to tell the difference between reality and the virtual world. Immersion is ``the dynamic interplay between visual, acoustic, and tactile feedback and the actions of looking around and manipulating objects" \cite{gibson2014ecological}. The degree of immersion is an objective property of the system and therefore measurable and quantifiable \cite{slater1999measuring} while presence is the user's response to an immersive environment. In our VR system, the avatar mediates the relationship between the user and the system. It is an extension and even a relocation of the user's body into virtual space, a vehicle through which the user is given embodied agency and presence in the virtual environment \cite{sageng2012philosophy}.

\subsection{Navigation in VR}

Two primary approaches to the problem of mapping small physical spaces to large virtual spaces for navigation and maneuvering have been explored in research: 
\begin{enumerate}
\item Locomotion: walking in place (natural \cite{usoh1999walking}, powered shoes or other devices \cite{iwata2006powered,iwata2007string,iwata2005circulafloor}), walking in space (natural \cite{welch1999hiball}, mechanical setups \cite{hale2014handbook}, redirection techniques \cite{williams2007exploring,steinicke2008taxonomy}), gestural walking \cite{nilsson2013perceived}
\item Abstractions or metaphors \cite{hale2014handbook}: miniaturized worlds, flying, driving, bicycling, teleporting, and virtual arm growing \cite{poupyrev1996go}. 
\end{enumerate}

We build upon the idea of natural locomotion in space allowing users to explore the virtual world by walking in the physical world. Not only do we track the user's position, we also track their full body as represented by 25 joints. This allows us to transfer the user's bodily motions to the onscreen avatar where a step taken forward in the real world is visible as an animation of a step taken forward by their avatar in the virtual world. A common limitation of a room-scale walking system is the difference in size of the physical and virtual worlds. A few mechanisms to overcome that difference have been explored. For example, a system that can imperceptibly rotate the virtual scene about the user \cite{kohli2005combining,razzaque2001redirected}, and a system that can apply redirection to steer users away from physical boundaries as they explore virtual environments \cite{suma2015making}. We are currently working on our own solution to allow exploration of large virtual spaces while walking in cluttered physical environments though that is not the focus of this paper.

\subsection{RGB-D Sensing}

Though RGB-D sensing devices have been custom-built for years, it is the computer gaming and home entertainment applications that have made them available for research outside specialized computer vision groups. The quality of the depth sensing, given the low-cost and real-time nature of devices like the Kinect, is compelling, and has made the sensor popular with researchers and enthusiasts alike. Available 3D reconstruction systems like KinectFusion \cite{izadi2011kinectfusion} enable a user holding and moving a standard Kinect camera to rapidly create detailed 3D reconstructions of an indoor scene. Reconstructing geometry using active sensors \cite{levoy2000digital}, passive cameras \cite{hartley2003multiple,merrell2007real}, online images \cite{frahm2010building}, or from unordered 3D points \cite{kazhdan2006poisson} are well-studied areas of research in computer graphics and vision. There is also extensive literature within the AR and robotics community on Simultaneous Localization and Mapping (SLAM), aimed at tracking a user or robot while creating a map of the surrounding physical environment \cite{thrun2002robotic}. Given this broad topic, and our need for building a VR environment that maps 1:1 to the physical environment, we used an existing reconstruction algorithm for 3D scanning. 

\begin{figure}[!t]
\centering
\includegraphics[width=0.9\columnwidth, height=0.8\columnwidth, keepaspectratio]{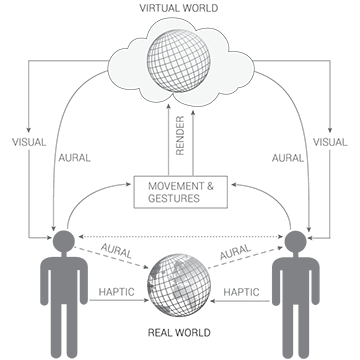}
\caption{Depiction of input and output pathways in our VR system. Users get visual feedback from the virtual world, audio feedback from both the real and virtual world, haptic feedback from the real world. Users use movement and gestures as input techniques and use voice to communicate.  }
\label{fig:setup}
\end{figure}

\section{MetaSpace II System Design}

We created three different VR scenes experienced using two different space and device configurations (see Figure 9). The first two involved scanning and mapping two different hallways to create scenes that matched each of those two physical spaces (see Figure 3a and 3b). In the third scene, we mapped an open lab space to a floating island virtual world (see Figure 3c and 3d). In the hallway/bridge scenarios, users were tasked with walking on the bridge toward each other and meeting midway for a high five. In the floating island scenario, users were tasked with grabbing a virtual cube and placing it on top of another virtual cube in order to unlock the portal by manipulating their physical box counterparts (see Figure 3d).

\begin{figure}[!t]
\centering
\includegraphics[width=0.9\columnwidth]{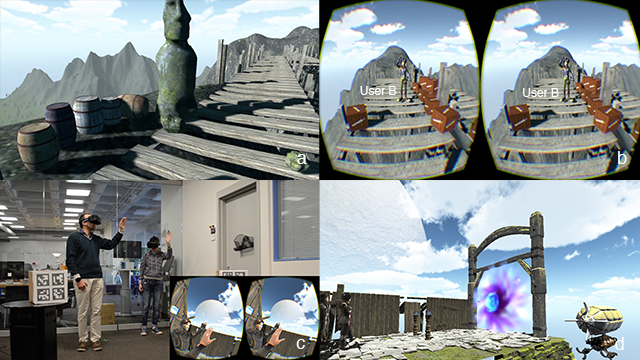}
\caption{We created three different VR scenes. a) A hallway represented as a virtual bridge with planters and pillars shown as barrels and a stone sculpture (see Figure 4). b) A different hallway also represented as a bridge with the drop off corresponding to a physical wall and the crates corresponding to storage cabinets in the real world. The view is seen from user B's perspective with user A waving at them. User A is shown in Figure 5 walking towards user B. c) Open lab space with two users in the VR scene shown in Figure 3d. Inset shows the view through the left users' HMD with the user on the right waving to them. d) Floating island scene that users in Figure 3d are immersed in. }
\label{fig:bridge}
\end{figure}

For each scene, we did a 3D scan of the real world space and used the resulting 3D model as a base on which to build the virtual world. Scanning gave us the correct relative sizes and positions of objects which is necessary to allow users to walk without colliding with walls or furniture as well as for interacting with physical objects. We could have achieved the same goal by measuring the physical space, furniture, and objects and manually creating a VR world in 3D modeling software. However, we envisioned an end to end system that comprised a virtual world creation mechanic with an output VR scene that users could experience, which made us choose the former method of building our VR world.

In the VR world, users are represented with full-body avatars and are tracked in real time such that they can walk around in the real world for exploring the virtual world. Users can interact with each other as well as with tracked physical objects and all their interactions are mirrored in the virtual world. Flow of data between two people using our system is depicted in Figure 2. We use a pair of Oculus Rift DK2 head-mounted displays (HMDs) to provide each user with an egocentric view of the virtual world, and Kinect (RGB-D) devices to capture user movements and to track the positions and orientations of objects.  We chose the Kinect camera as it is widely accessible and low cost compared to virtual reality lab setups that usually have expensive OptiTrack or Vicon tracking systems.

\subsection{Pipeline}

A depth sensing device (Google Tango tablet) was used to scan the hallways and lab space (see Figure 4) using an existing 3D scanning application from the Google Play store. In our lab space, one wall is entirely made of glass. Because sensing glass with IR does not work, we manually added a wooden fence to our scene at the location of the real world glass wall.  The initial scan gave us a mesh with a very large number of vertices and triangles and some holes. We imported the mesh into MeshLab to reduce triangle count and output a watertight mesh. The output from MeshLab was imported into Blender, where we prepared the models for import into Unity3D for building the VR scenes. Unity3D is a cross platform game engine with support for networking, cloud-based rendering, texture mapping and lighting to name a few.

\begin{figure}[!t]
\centering
\includegraphics[width=0.9\columnwidth]{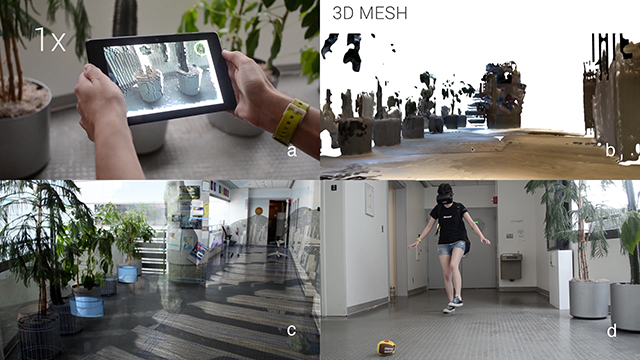}
\caption{The entire process from scanning to exploring the VR world. a) 3D scanning of the real world space using the Google Tango tablet. b) A mesh is generated from the scanned point cloud data. c) The generated mesh is used as a base for creating the VR scene in Unity3D where virtual objects (barrels, sculpture, bridge) corresponds to the real world (planters, pillar, hallway). d) A user about to start walking down the hallway/bridge while wearing an Oculus Rift DK2.}
\label{fig:setup}
\end{figure}

We use Kinect devices to track user skeletons. In skeletal tracking, a human body is represented by a number of joints (25 for Kinect V2) that represent body parts such as head, neck, shoulders, and arms. Each joint is represented by its 3D coordinates and mapped to the joints in a rigged 3D model. Rigging is the process of creating bones and joints for the avatars and allows the animation system to pose and manipulate them as needed. Joint data from the Kinect is used to update the movements of the rigged models in Unity3D after mapping joing data for each skeleton to bone data for each avatar. This is important as we want to animate the character's movements using the real world movements of the users. We use RGB data from the Kinects and the OpenCV library to track position and orientation of two boxes. Each physical box is represented as a virtual cube in our scene and appears at at corresponding location in VR to it's physical location. The positions and orientations of the virtual cubes match those of the physical boxes.

\subsection{Hardware and Software}

Our server is implemented using C++ and OpenCV. Our clients are implemented using Unity and the Oculus Rift plugin. The server is an Intel Core i5-4200 2.3GHz Windows laptop with 4GB RAM. One client is an Intel Core i7-3520 2.9GHz Windows laptop with 8GB RAM. The second client is a dual boot Intel Core i5-4258U 2.6GHz OSX laptop with 16GB RAM. 

\section{Key Components}

Our goal is to leverage the physicality of the real environment and use our first person experiences with the real world for enhancing the sense of presence in virtual first person experiences. We had about 40 users try out our system in April 2015 and another 50-60 try a different version in October/November 2015. Users walked, waved, high five'd, grabbed a box and played with it, and stacked boxes. We received positive feedback about the full-body avatar and the object interaction, as it appeared users were most excited about those elements. Though these were not formal user studies, we are very excited about conducting one to understand how a user's sense of presence in VR is affected by the elements included in our system.

\subsection{Body Tracking}

\begin{figure}[!t]
\centering
\includegraphics[width=0.9\columnwidth]{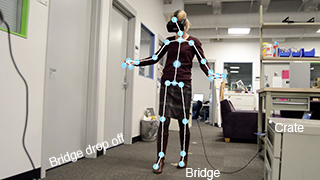}
\caption{User A walking down the hallway (real) and bridge (virtual) towards User B (see Figure 3b). Both users' skeletons are tracked by the Kinect and their physical movements are mapped to their avatar's movements.}
\label{fig:bridge}
\end{figure}

In order for each user to have agency in the VR environment, we track them in real-time using Kinect devices and it's skeleton tracking system. Each joint is represented by its 3D coordinates and mapped to the joints of a rigged 3D model. Thus each user dynamically controls an avatar through their body movements. This allows users to walk around in physical space, for example, a hallway, while walking on a bridge in virtual space (see Figures 3 and 5). 

Real-walking in virtual environments is more natural and produces a higher sense of presence than other locomotion techniques \cite{usoh1999walking,slater1995taking}. However, this is typically restricted in size to the area of tracked space. Reorientation techniques have traditionally been utilized to enable free exploration of infinitely large VR spaces without the use of joysticks, walking-in-place techniques, omni-directional treadmills or bicylces \cite{brooks1987walkthrough,darken1997omni,iwata1999walking}. Though our current system provides a room-scale VR experience, we are working on the next version where we remove that limitation by employing an inside-out position tracking system paired with motion capture systems for full body tracking to allow us to expand our room-scale VR to an entire building floor and beyond.

\subsection{Object Tracking}
Our approach for interacting with physical objects builds on the concept of passive haptics \cite{hoffman1998physically}, i.e. receiving feedback from touching a physical object that is registered to a virtual object. The physical properties of the tangible interface can be used to suggest ways in which the attached virtual objects may be used, in essence adding affordances to virtual items. Object tracking is an component in a wide range of applications in computer vision, such as traffic monitoring, video indexing, and self driving vehicles. Given the initialized state (e.g., position and size) of our boxes in an RGB frame from the Kinect, the goal of tracking was to estimate the states of the boxes in the subsequent frames and applying the learned transformations to their virtual counterparts in our VR scene. Although object tracking has been studied for several decades, and much progress has been made \cite{yilmaz2006object}, it remains a very challenging problem, especially for real-time application scenarios. We chose a marker based instead of a natural feature-based 3D object tracking system due to speed of implementation and higher frame rates as determined empirically for our scenario. Numerous factors affect the performance of a tracking algorithm, such as illumination variation, occlusion, as well as background clutter, and there exists no single tracking approach that can successfully handle all scenarios.

\subsection{Texturing}

Previous research has focused on virtual objects being modeled on the physical proxy \cite{hoffman1998physically}. However, our goal is not the reconstruction of a realistic virtual replica of the physical space or objects. We want to use the physical space as a template for placement of walls and furniture in the virtual world such that a correspondence between the two spaces exists and are users are able to walk freely in the real world while wearing HMDs. We do want to change the visual appearance of virtual objects by texturing the 3D models differently from their real world textures. For example, the walls in our scene corresponding to the physical lab space look like wooden fences (see Figure 3d). They can be made to look like stone walls of a cave or those of a space station. Similarly, furniture can change appearance and chairs and tables can look like they belong to a Victorian living room. 

\begin{figure}[!t]
\centering
\includegraphics[width=0.9\columnwidth]{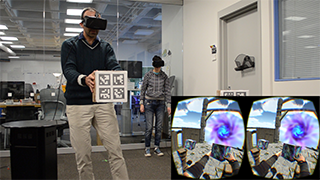}
\caption{A user grabs a box in the real world by reaching out and picking up the cube in the VR world. Their task is to stack the two boxes/cubes to unlock the portal.}
\label{fig:bridge}
\end{figure}

We aim to match the virtual representation closely with the physical characteristics of the actual objects in order to maintain sensory coherence. For example, an empty cardboard box feels light, non-slippery, a little rough, and has a distinct sound when touched or moved, and so on. Our virtual cube therefore needs to match some of these tactile characteristics to maintain coherence in sensory feedback. The texture of the virtual box should be such that it appears light enough to be picked up and the material on it is not metal or soft foam as they would completely alter the user's expectations of how it would feel and create a mismatch with what it actually feels like when they pick up the real box/virtual cube (see Figure 6). Additionally, there needs to be an approximate size match or the user may not be able to pick up the real box without some trial and error, due to proprioception. Objects that present similar affordances \cite{gibson1977theory} in the parts most likely to be interacted with, are the best candidates for virtualizing and using as passive haptics.  

\subsection{Feedback}

The Reality-Virtuality continuum \cite{wiki:reality-virtuality} defines a range of possible ``realities" affecting our senses in different ways.  At one end is the Real Environment and at the other end is the Virtual Environment, encompassing all possible variations and compositions of real and virtual objects. In between is the notion of Mixed Reality (MR), encompassing Augmented Reality (AR)  and Augmented Virtuality \cite{milgram1994taxonomy}. These types of systems are usually characterized by adding or subtracting elements of either side of the continuum to the user experience.  Our system spans the continuum as the user interacts with real world objects and walks in the real world while interacting with a fully virtual environment. To address the difficulty in rendering convincing physical and haptic experiences in VR, we explore the concept of passive haptics. 

VR systems and devices are now able to create visually and aurally convincing experiences. Current technology however, is still not able to stimulate our other senses with the same resolution. In particular, even though some technologies provide the experience of tactile feedback (e.g., haptic
devices), force feedback (e.g. exoskeletons) and locomotion (e.g., omni-directional treadmills), these devices are not able to mimic how we experience the real world. For example, haptic devices predominantly use vibrotactile feedback \cite{hale2014handbook} which is just one type of tactile feedback that we may encounter in the real world. Touch also conveys information such as contact surface geometry, roughness, slippage, and temperature. Through passive haptics, our system is able to convey realistic haptic feedback to the user (see Figure 6). Additionally, we receive force feedback \cite{bouzit2002rutgers} in the real world which provides information on object surface compliance, weight, and inertia. While treadmills allow users to navigate endlessly large spaces, they are unable to render physical obstacles or other terrain properties, such as ridges and uneven elevations. 

\subsection{Networking}

For realtime performance, we adopt a client-server, distributed architecture to address computational complexity involved in object tracking. For a synchronous multiuser virtual experience, it is essential that all persons receive the same exact system state at all times. In Figure 7a, we show our client-server architecture that is made of four main entities: cloud server providing middleware and service (Photon Server), laptops running local clients (Unity3D), Kinect RGB-D tracking devices, and HMDs (Oculus Rift DK2s). The Kinect devices track user movement and that data is sent to each client through the cloud server such that each client is synced and displays the same system state to each user. In Figure 7b, we show a second client-server architecture that is made of three main entities: user and object tracking server (C++), laptops running local clients (Unity3D), and HMDs (Oculus Rift DK2s).  Data is read in from the Kinect and through the cloud server sent to all clients. Joint orientations are calculated from the joint position data received from the server and applied to the avatars. The virtual cubes receive the tracked object pose which transforms their position and orientation on the client.

\begin{figure}[!t]
\centering
\includegraphics[width=1.0\columnwidth]{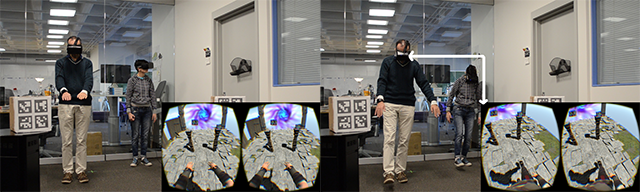}
\caption{A user checks out their hands and body in the VR scene before walking up to interact with the box/cube.}
\label{fig:bridge}
\end{figure}

\begin{figure}[!t]
\centering
\includegraphics[width=0.9\columnwidth]{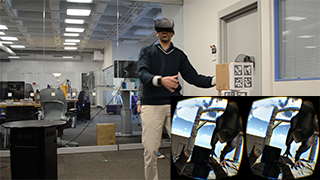}
\caption{The instant boxes/cubes are stacked, the spiral vortex blocking the portal disappears and a T-Rex rushes out towards the adventurers.}
\label{fig:bridge}
\end{figure}

\section{User Experience}

This section is about the user experience in the VR scene that corresponds to the physical setup shown in Figure 9b. The two adventurers try to make sense of their surroundings: their airship, seemingly grounded beyond the shimmering air. Where are they? The moon rises massively in the sky as if only a short distance away. Floating above a wooden fence to the left, its fullness is obscured by clouds so close they could reach out and touch them. They realize, no, their airship is not grounded, they have been stranded on some floating island. Floating over what, they cannot tell. Beyond the edges of this land mass they would fall into an infinity of space.  A swirling energy field appears before them. The one point of access to something outside of this island is their ship. They determine the key to their escape lies within this portal. But how can they open this portal to allow them their freedom. They notice some relationship with portal and the two pulsating cubes resting on pedestals before them. Somehow they need to interact with these cubes to be able to gain access to what secrets or escape the portal can give. Will it allow them to board their ship or will it take them somewhere entirely new. And so their adventure begins. Little do they know that opening the portal will let loose a T-Rex (see Figure 8).

Experiencing MS2 consists of three parts. First is receiving visual input through the Oculus Rift Dk2 head mounted display devices, one for each person. Second is being able to fully control a virtual avatar through bodily movements in the real world, and third is interacting with objects in the VR scene and receiving haptic feedback. We asked both users to start by looking around the VR scene and looking at their hands, feet, and body (see Figure 7) followed by acknowledging each others presence in the virtual world. Each user was then asked to walk up to the virtual cube on the pillar a few feet in front of them, to pick it up and put it on top of the other cube. Communication between the users and the virtual world happens through physical movement tracked using the Kinect depth sensor. Interaction with objects happens through physically grasping and moving them as they are tracking using the Kinect RGB sensor. These mechanisms help create a natural immersive multiuser experience where interacting with virtual objects is done through interacting with their physical counterparts.

\begin{figure}[!t]
\centering
\includegraphics[width=0.8\columnwidth]{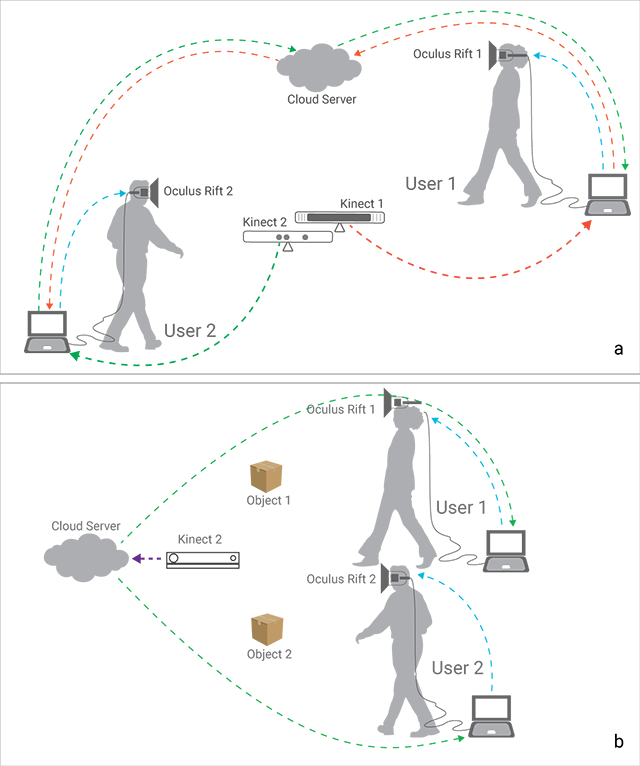}
\caption{MetaSpace system configurations. Each user wears an Oculus Rift DK2 HMD and can see a first person view of the virtual world and see other user's who are also present in the same virtual world with them.  }
\label{fig:setup}
\end{figure}

Digital games often present players with tasks where they need to use objects such as swords, guns, lanterns etc. These objects are unlikely to be found in users' living rooms but that does not need to be a limitation impacting the types of experiences that can be designed in our VR system. In several instances, we believe the role for passive haptics would almost be like that of toys used in children's imaginative play where a flashlight becomes a light saber or a pencil turns into a magic wand. Because of this, we think an exact match between real and virtual counterparts would not be required to successfully provide haptic feedback. Similarly shaped and sized objects easily found around the house could fulfill the role successfully and provide the necessary feedback to create a believable VR experience. Every element in the VR scene does not need a physical counterpart either. The dinosaur does not have any physical counterpart, but it's fantastical elements mixed with real objects and environments that are part of the lure of VR, of the escape into new worlds. 

\section{Conclusion}

We have presented the design and realization of a social and immersive virtual reality system and described the process of how we created our virtual worlds. By employing RGB-D devices, head-mounted displays, and a client-server architecture, multiple users simultaneously inhabit the virtual world and interact with each other as well as with the real world through VR. We discussed the implementation details and optimization strategies of each component and their role in creating an overall VR experience. Our system runs in real-time and is indeed a proof of concept for practical applications based on social VR systems with haptic feedback and full-body tracking. Since our system has already been tried by over a 100 people, our next plan is to conduct a formal evaluation of users' experiences for a variety of tasks involving interaction with people and with objects. 


\bibliographystyle{acm-sigchi}
\bibliography{metaspace2}
\end{document}